\documentclass[11pt,dvips]{article}
cm \voffset = -2truecm

\usepackage{graphicx}
\usepackage{amssymb}
\usepackage{latexsym}

\begin{document}
\begin{titlepage}
\setcounter{page}{1}
\renewcommand{\thefootnote}{\fnsymbol{footnote}}

\vspace{5mm}
\begin{center}

 {\Large \bf
 Bipartite entanglement of multipartite coherent states using quantum network of beam splitters}

\vspace{1cm}

{\bf M. Daoud$^{a,b}$\footnote {daoud@pks.mpg.de,
m$_-$daoud@hotmail.com}} and {\bf E. B.
Choubabi$^{c}$\footnote{choubabi@gmail.com}}

% and {\bf Ahmed
%Jellal$^{c,d,e}$\footnote{jellal@pks.mpg.de, jellal@ucd.ma}}

\vspace{5mm}

{$^{a}$\em  Max Planck Institute for Physics of Complex Systems,
N\"othnitzer Str. 38,\\
 D-01187 Dresden, Germany}\\

{$^{b}$\em Department of Physics,
Faculty of Sciences, Ibn Zohr University},\\
{\em PO Box 8106, 80006 Agadir, Morocco}

{$^{c}$\em Theoretical Physics Group,  %Department of Physics,
Faculty of Sciences, Choua\"ib Doukkali University},\\
{\em PO Box 20, 24000 El Jadida, Morocco}

%{$^d$\em Physics Department, College of Sciences, King Faisal University,\\
%PO Box 380, Alahsa 31982, Saudi Arabia}

%{$^e$\em Saudi Center for Theoretical Physics, Dhahran, Saudi Arabia}

\vspace{3cm}

\begin{abstract}

We investigate the generation of multipartite entangled $SU(k+1)$
coherent states using a quantum network involving a sequence of $k$
beam splitters. We particularly investigate the entanglement in
multipartite $SU(2)$ coherent states ($ k = 1$). We employ the
concurrence as measure of the degree of bipartite entanglement.

\end{abstract}
\end{center}
\end{titlepage}

\newpage

\section{Introduction}

Due to seminal works on quantum teleportation \cite{Ben1},
superdense coding \cite{Ben2} and quantum key distribution
\cite{Eckert}, entanglement in multipartite quantum systems began to
be recognized as valuable resource for performing communication and
computational tasks \cite{Fuchs,Rausschendorf,Gottesman}. This
motivated  the considerable interest in the development of a
quantitative theory of entanglement and the definition of its basic
measure (concurrence, entanglement of formation and linear entropy
\cite{Rungta,Ben3,Wootters,Coffman}). The physical implementation of
entangled states is then of major importance and several different
physical systems have been considered. We mention the generation of
entangled electromagnetic states  using type I or type II parametric
down conversion \cite{Kwiat}. Another experimentally accessible
device which can be used to generate optical  entangled states is
the beam splitter \cite{Tan,Sanders1,Sanders2,Paris}. In quantum
optics  the action of a beam splitter, which is essentially a mixer
of two electromagnetic modes, can be represented by a unitary
operator relating the input and the output states. In general, the
output state is a superposition of the Fock states which is
entangled, except the harmonic oscillator coherent states who do not
exhibit entanglement when passed through one arm of 50:50 beam
splitter while the second arm is left in the empty vacuum state
\cite{Kim}. The beam splitter device provides a very useful
technique to investigate the entanglement of nonorthogonal states as
for instance the entangled coherent states introduced in
\cite{Sanders1,Sanders2,Mann,Wang,Fu}. They found a notable success
in the context of quantum teleportation \cite{van,Jeong1}, quantum
information processing \cite{Jeong2,Yang}, tests of local realism
\cite{Sanders1} and very recently an approach for quantum repeaters
with entangled coherent states was proposed in \cite{Sangouard}. All
those applications explain their intensive investigation. In this
sense, the entanglement behavior of $SU(2)$ spin coherent states,
when passed through a beam splitter, has been previously considered
in \cite{Markham}. Similar study was done in \cite{Gerry1} for
$SU(1,1)$ coherent states defined in an infinite
 dimensional Hilbert space.  Entangled quantum systems can exhibit correlations
that cannot be explained on the basis of classical laws and the lack
of entanglement
 in a collection of states is clearly a signature of classicality \cite{Markham}.
The fact that the coherent states minimize the quantum fluctuations
\cite{Klauder,Perelomov} and subsequently present  semi-classical
behavior, constitutes the main motivation to use them for the
understanding of bipartite
\cite{Sanders1,Sanders2,Mann,Wang,Fu,Gerry2,Luis} as well as the
multipartite entanglement of harmonic oscillator  coherent states
\cite{Wang1,Wang2}.

In this paper, we shall be interested in the bipartite entanglement
of generalized coherent states generated from the quantum
electromagnetic states passing through a quantum network of $k$ beam
splitters. We show that under some special  assumptions, one can
generate $SU(2)$ for $k =2$, $SU(3)$ for $k=3$ or more generally
$SU(k+1)$ multipartite coherent states which are labeled by
continuous variables related to reflection-transmission coefficients
of the beam splitters.

The outline of the paper is as follows. In section 2, we define the
action of a quantum network of $k$ beam splitters. We show that this
action leads $SU(k+1)$ coherent states, in the Perelomov sense,
labeled by complex variables related to reflection and transmission
parameters of beam splitters. This means that multi-port beam
splitters are applicable in optical realizations of coherent states
associated with higher symmetries. In section 3, we consider the
superpositions of equally weighted $SU(2)$ multipartite coherent
states arising from two quantum networks of beam splitters
characterized by different reflection-transmission parameters. By
equally weighted, we mean that each multipartite wave function  is
equally balanced with the other element in the superposition. We
divide the entire system into two subsystems to  discuss the
bipartite entanglement. We employ the concurrence \cite{Hill} to
determine the degree of entanglement. We consider the pure and mixed
state cases.
 Concluding remarks close
this paper.

\section{ Quantum network of beam splitters}

As mentioned above, the study of entangled states in the last decade
has revived interest in the beam splitter because this device offers
us an simple way to probe the quantum nature of electromagnetic
field by means of simple experiments. We recall that the beam
splitter is an optical element with two input ports and two output
ports which, in some sense, governs the interaction of two harmonic
oscillators. The input and output boson operators are related by a
unitary transformation which may be viewed as an element of the
$SU(2)$ group. Recently, a quantum network of beams splitters was
used to create multi-particle entangled states of continuous
variables \cite{Van} and also multi-particle entangled coherent
states \cite{Wang2} associated with the usual harmonic oscillator
algebra.

\subsection{ Fock states output}

We shall First discuss a simplest beam splitters combination that
enables the generation of coherent states associated with the Lie
algebra $su(k+1)$ using $k+1$ different electromagnetic field modes
($ k \in {\mathbb N}-\{0\}$). To this end, we define a sequence of
$k$ beam splitters ("$k$-beam splitters") acting as follows. The
output of the  first beam splitter, which is a superposition of the
modes $0$ and $1$, is injected into the first arm of the second beam
splitter while the second arm receives modes of quantum
electromagnetic field labeled as $2$. The output state of the second
beam splitter is then injected in the the first port of the third
beam spitter while the other port receives the photons in the mode
3. This scheme can be generalized to an arbitrary number of beam
splitters such that the output state of the each beam splitter, of
the sequence under consideration, is inserted in one port of the
next beam splitter while another electromagnetic mode is inserted
into the other input port. The figure 1 give the schematics of this
combination of beam splitters.
\begin{center}
\includegraphics[width=4in]{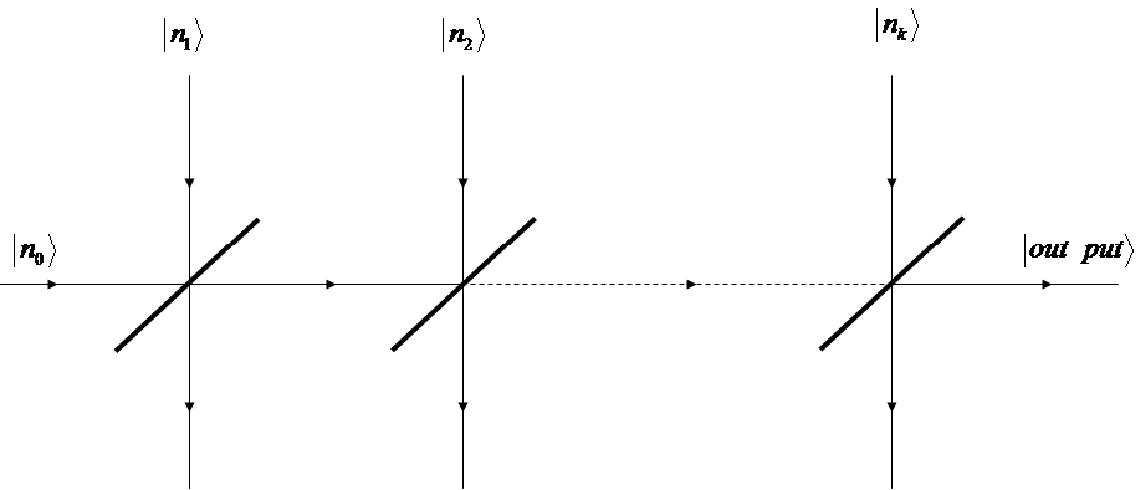}\\
Fig. 1: {\sf The sequence  of $k$ beam splitters.}
\end{center}
Then the unitary transformation to realize this can be expressed as
the product of a sequence of beam splitter tansformations. It is
given by
\begin{equation}
{\cal U}_k = {\cal B}_{k-1,k}(\theta_k)  {\cal
B}_{k-2,k-1}(\theta_{k-1}) \cdots {\cal B}_{0,1}(\theta_{1}),
\label{operatorB}
\end{equation}
 where  the operators
\begin{equation}
{\cal B}_{l,l+1}(\theta_{l+1}) = \exp(\frac{i}{2} \theta_{l+1}
(a_l^+a_{l+1}^- + a_l^-a_{l+1}^+)) \label{operatorBl}
\end{equation}
are the ordinary $SU(2)$  beam splitters for $(l = 0, 1 , 2, \cdots,
k-1)$ and the angles $\theta_l$ are related to  reflection and
transmission coefficients as follows
\begin{equation}
 t_l = \cos\frac{\theta_l}{2} \qquad  r_l = \sin\frac{\theta_l}{2}.
\end{equation}
In the equation (\ref{operatorBl}), $a^+_l$ and $a^-_l$ stand  the
creation and annihilation operators of $k+1$ independent bosonic
oscillator. They satisfy the usual commutation rules
\begin{equation}
 [a^-_l , a^+_{l'}] = \delta_{ll'} , \qquad  [a^-_l , a^-_{l'} ] =   [a^+_l , a^+_{l'} ] = 0, \qquad  l,l' = 0, 1, 2, \cdots k.
\end{equation}
As Hilbert space let chose the infinite dimensional  Fock space
generated by  multi-mode states
\begin{equation}
{\cal H} =  {\cal H}_0 \otimes {\cal H}_1 \cdots \otimes {\cal
H}_{k} = \{ \vert n_0 , n_1, \cdots , n_{k} \rangle ,  n_i \in
\mathbb{N}\}.
\end{equation}
If the input state is given by $\vert n_0 , n_1, \cdots , n_{k}
\rangle$, the action of the operator  ${\cal U}_k $ is given by the
following Fock states superposition
\begin{equation}
{\cal U}_k \vert n_0 , n_1, \cdots , n_{k} \rangle =
\sum_{m_0,m_1,\cdots,m_{k}} C_{n_0 , n_1, \cdots , n_{k}
}^{m_0,m_1,\cdots,m_{k}}
 \vert m_0 , m_1, \cdots , m_{k} \rangle
\end{equation}
where the coefficients $C$ stand for the matrix elements of the
unitary operator ${\cal U}_k$. In general the output is a
$(k+1)$-particle entangled state. On the other hand, the action of
the unitary operator ${\cal U}_k$ on the state $\vert n_0 = n , 0,
\cdots , 0 \rangle$ gives
\begin{equation}
{\cal U}_k \vert n , 0, \cdots , 0 \rangle = {\cal C}
\sum_{n_1=0}^{n}\sum_{n_2=0}^{n_1} \cdots \sum_{n_{k}=0}^{n_{k-1}}
\frac{ \xi_{1}^{n_1}\xi_{2}^{n_2}\cdots
\xi_{k}^{n_{k}}\sqrt{n!}}{\sqrt{(n -
n_1)!(n_1-n_2)!\cdots(n_{k-1}-n_{k})!n_{k}!}} \vert n-n_1 , n_1-n_2,
\cdots , n_{k} \rangle \label{suncs}
\end{equation}
where the normalization constant is given by
\begin{equation}
 {\cal C}  =  (1 +  |\xi_{1}|^2 + |\xi_{1}|^2|\xi_{2}|^2\cdots +|\xi_{1}|^2|\xi_{2}|^2\cdots |\xi_{k}|^2)^{-\frac{n}{2}},
\end{equation}
and the new variables $\xi$ are defined by
\begin{equation}
\xi_l = i t_{l+1}~\frac{r_l}{t_l}\qquad {\rm for} \qquad l =
1,2,\cdots,k-1 \qquad {\rm and} \qquad \xi_{k} = i
\frac{r_{k}}{t_{k}}
\end{equation}
in terms of the  reflection and transmission coefficients of the
beam splitters constituting the network. Then, the output state
(\ref{suncs}) turns out to be the $SU(k+1)$ coherent state
associated with the completely symmetric representation labeled by
the integer $n$. Indeed, the bilinear product of creation and
annihilation modes realizes, \`a la Schwinger, the generators of the
Lie algebra $su(k+1)$ and the operator ${\cal U}_k$ is nothing but a
unitary displacement operator which acting on the highest weight
vector $\vert n_0=n, 0, \cdots, 0\rangle$, gives the $SU(k+1)$
coherent states in the Perelomov sense (see for instance
\cite{Daoud1} for more details). This is clarified in the what
follows.

\subsection{$SU(k+1)$ coherent states}

In this subsection we discuss the relationship between the states
(\ref{suncs}) and the $SU(k+1)$ coherent states. To do this, it is
interesting to note that the states (\ref{suncs}) write as an
expansion of the number states belonging the the restricted Fock
space
$${\cal F} = \{ \vert n_0,n_1 , n_2 \cdots , n_k \rangle, n_0+n_1+n_2 +\cdots+n_k = n\}$$
which is finite dimensional. In order to show that (\ref{suncs}) are
$SU(k+1)$ coherent states, we  realize the $su(k+1)$ generators as
bilinear products in creation and annihilation operators. They are
given by
$$ e_i^- = a_i^+a_{i+1}^- \quad   e_i^+ = a_i^-a_{i+1}^+  \quad i = 0, 1,\cdots, k,$$
and the Cartan generators
$$h_i = a_i^+a_i^- - a_{i+1}^+a_{i+1}^-, \quad i = 0, 1, \cdots, k-1.$$
The generators $ e_i^-$ and $ e_i^+$ are the so-called Weyl
operators. The generators of $su(k+1)$ with a non trivial action
(non vanishing and non diagonal) on the fiducial vector (or the
highest weight vector of the symmetric representation) are
$$t_i^+ = a_0^-a_i^+   \qquad t_i^- = a_0^+a_i^- \qquad i = 1, 2,\cdots, k.$$
They can be defined from the $su(k+1)$ generators as follows
$$ t_1^+ = e_0^+, \qquad t_{i+1}^+ = [ e_i^+ , t_{i}^+],$$
$$ t_1^- = e_0^-, \qquad t_{i+1}^- = [ t_{i}^- , e_i^- ],$$
for $i = 1, 2, \cdots , k-1.$ Using the raising and lowering
operators $t_i^+$ and $t_i^-$, the states (\ref{suncs}) can be shown
to have the displacement forms. Namely, they are created by the
action of the unitary displacement operators in $SU(k+1)$ acting on
the state $\vert n=n_0, 0, \cdots , 0\rangle$ (the vacuum). Indeed,
it is simply verified that the action of the unitary operator
$$D(\xi_1 , \xi_2, \cdots, \xi_k) = \exp(\sum_{i=1}^{k} (\xi_i t_i^+  - \bar \xi_i t_i^-))$$
on the state $\vert n=n_0, 0, \cdots , 0\rangle$ gives the states
(\ref{suncs}). This is exactly the Perelomov definition of coherent
states for Lie algebras. Thus, it is clear that the output states
(\ref{suncs}) constitutes a special class of $SU(k+1)$ coherent
states.

To close this subsection, it is important to stress that the
generation of $SU(k+1)$ coherent states using the optical chain of
$k$ beam splitters described above requires input radiation state
with fixed number of photons. The experimental production of such
interesting and highly non classical states has been investigated
during the last decade (see \cite{Brattke} and references therein).
Recently an important experimental advance was reported by Hofheinz
et al in \cite{Hofheinz}. They gave the first experimental
demonstration for generating photon number Fock states containing up
to $n = 6$ photons in a super-conducting quantum circuit.

\subsection{ Contraction of $su(k+1)$ algebra and Glauber coherent states}
At this stage, it is natural to ask about the relation between
$SU(k+1)$ coherent states obtained above and the standard Glauber
coherent for multi-mode electromagnetic fields. In this subsection,
we shall show that he usual Glauber coherent states can be obtained
from the $SU(k+1)$ as a special limiting case. To show this, we
rewrite the coherent states (\ref{suncs}) as
\begin{equation}
 \vert \zeta_{1} , \zeta_{2}, \cdots , \zeta_{k} \rangle = {\cal C}
\sum_{n_1=0}^{n}\sum_{n_2=0}^{n-n_1} \cdots \sum_{n_k=0}^{n -
\sum_{i=1}^{k} n_i} \frac{ \zeta_{1}^{n_1}\zeta_{2}^{n_2}\cdots
\zeta_{k}^{n_{k}}\sqrt{n!}}{\sqrt{(n - \sum_{i=1}^{k}
n_i})!n_1!n_2!\cdots n_{k}!} \vert n -\sum_{i=1}^{k}n_i , n_1, n_2,
\cdots , n_{k} \rangle \label{zeta}
\end{equation}
where the new variables are defined by
$$\zeta_i = \xi_1 \xi_2 \cdots \xi_i \qquad i = 1, 2, \cdots, k.$$
 Let $n \to \infty $, $\vert
\zeta_i \vert \to 0$ in such a way that the product $ n \vert
\zeta_i \vert^2 = \vert z_i \vert^2$ is fixed. In this limit, the
coherent state (\ref{zeta}) tends to
\begin{equation}
\vert \zeta_1 , \zeta_2, \cdots , \zeta_k \rangle \to \vert n
\rangle \otimes \vert z_1 \rangle  \otimes \vert z_2 \rangle \cdots
\otimes \vert z_k \rangle \label{csho}
\end{equation}
where
$$\vert z_i \rangle = \exp(-\vert z_i \vert^2/2)\sum_{n_i=0}^{\infty} \frac{z_i^{n_i}}{\sqrt{n_i!}} \vert n_i \rangle$$
are the usual Glauber coherent states. This limit can understood as
contraction of $su(k+1)$ algebra into $k$ independent
Weyl--Heisenberg algebras. Indeed, using
$$ t_i^+ \vert n_0, n_1, \cdots n_i, \cdots, n_k \rangle = \sqrt{n_0(n_i+1)} ~\vert n_0, n_1, \cdots n_i+1, \cdots, n_k \rangle $$
$$ t_i^- \vert n_0, n_1, \cdots n_i, \cdots, n_k \rangle = \sqrt{(n_0+1)n_i} ~\vert n_0+1, n_1, \cdots n_i-1, \cdots, n_k \rangle$$
and the relation $n_0 = n - (n_1+n_2+ \cdots+n_k)$ satisfied by the
states of the $SU(k+1)$ invariant Fock space ${\cal F}$, one can
simply verify that for $n$ large
$$t_i^+ \sim \sqrt{n}~ a_i^+ \qquad t_i^- \sim \sqrt{n} ~a_i^-.$$
It follows that in this limiting case, we end up with $k$ commuting
copies of usual harmonic oscillator and  the states (\ref{zeta})
gives
$$\vert \zeta_1 , \zeta_2, \cdots , \zeta_k \rangle \to \exp(\sum_{i=1}^{k} (z_i a_i^+  - \bar z_i a_i^-))\vert n , 0 , 0, \cdots , 0 \rangle$$
where the right hand coincides with $\vert n \rangle \otimes \vert
z_1 \rangle  \otimes \vert z_2 \rangle \cdots \otimes \vert z_k
\rangle$ reflecting that, in this special case, the output state is
completely separable.

\subsection{ Generation of multipartite nonorthogonal states}

Here we provide an optical scheme to generate multipartite $SU(k+1)$
coherent states. It clear that with $p$ decoupled network of quantum
beam splitters, on can generate a tensorial product of $SU(k+1)$
coherent states. For instance, the special case $k = 1$
corresponding to a single beam splitter where $n$ photons are
injected into one port with only the vacuum at the other port,
generates $SU(2)$ coherent state. To generate a multipartite $SU(2)$
coherent states, we consider a $p$ uncoupled beam splitters acting
on the Hilbert space $ {\cal H} = {\cal H}_1 \otimes {\cal H}_2
\otimes \cdots \otimes {\cal H}_{2p}$. This  is a tensorial product
of $2p$ usual bosonic Fock space. The beam splitter ${\cal
B}_{i,i+1}(\theta_{i})$, with $ i = 1, 3, \cdots , 2p-1$, acts on
the Fock space ${\cal H}_i \otimes {\cal H}_{i+1}$ as coupler of the
modes $i$ and $i+1$ . The action of this set of uncoupled beam
splitters is described by the following transformation
\begin{equation}
{\cal T}_{p,1} = {\cal B}_{2p-1,2p}(\theta_{2p-1})  {\cal
B}_{2p-3,2p-2}(\theta_{2p-3}) \cdots {\cal B}_{3,4}(\theta_{3}){\cal
B}_{1,2}(\theta_{1}). \label{operatorBp}
\end{equation}
We assume that the input state of each beam splitter, characterized
by the operation ${\cal B}_{i,i+1}(\theta_{i})$, is $\vert n , 0
\rangle \in {\cal H}_i \otimes {\cal H}_{i+1}$ where $n \in
\mathbb{N}-\{0\} $ is an arbitrary photon number. Then, The input
state of the entire system is given
\begin{equation}
\vert {\rm input} \rangle = \vert n , 0, n, 0, \cdots, n,0\rangle
\label{input},
\end{equation}
and one can see that the output state
\begin{equation}
{\cal T}_{p,1} \vert {\rm input} \rangle  = \vert \alpha_1 \rangle
\otimes \vert \alpha_3 \rangle \otimes \cdots \otimes \vert
\alpha_{2p-1} \rangle \label{su2}
\end{equation}
where $\vert \alpha_i \rangle$ ($i = 1, 3, \cdots , 2p-1$) are the
$SU(2)$ coherent states given by
\begin{equation}
\vert \alpha_i \rangle = (1 +  |\alpha_{i}|^2
)^{-\frac{n}{2}}\sum_{m =0}^{n}
 \frac{\sqrt{n!}}{\sqrt{(n - m)!m!}} \alpha_{i}^{m} \vert n - m, m
 \rangle \label{alphacs}
\end{equation}
with the labeling parameter $\alpha_i = ir_i/t_i$ is expressed in
terms of the reflection-transmission rate.

In a similar way, one can extend this scheme to generate states
which are tensorial product of $SU(3)$ coherent states. This can be
done as follows. We consider set of $p$ decoupled chain of beam
splitters corresponding to $k = 2$. Each $k = 2$ chain of beam
splitters is described by the unitary transformation ${\cal
B}_{i+1,i+2}(\theta_{i+1})  {\cal B}_{i,i+1}(\theta_{i})$ acting on
the three particle Hilbert space ${\cal H}_i\otimes{\cal
H}_{i+1}\otimes{\cal H}_{i+2}$ where $i$ takes the integer values
$1, 4, 7, \cdots , 3p-2$. It follows that the entire system is
described by the following transformation
\begin{equation}
{\cal T}_{p,2} = {\cal B}_{3p-1,3p}(\theta_{3p-1})  {\cal
B}_{3p-2,3p-1}(\theta_{3p-2}) \cdots {\cal B}_{5,6}(\theta_{5}){\cal
B}_{4,5}(\theta_{4}) {\cal B}_{2,3}(\theta_{2}){\cal
B}_{1,2}(\theta_{1}).
\end{equation}
 The action of ${\cal T}_{p,2}$ on the
$3p$-particles states of type $\vert n, 0, 0, n, 0, 0, \cdots, n, 0,
0\rangle$ gives
\begin{equation}
{\cal T}_{p,2} \vert n, 0, 0, n, 0, 0, \cdots, n, 0, 0\rangle =
\vert \beta_1, \beta_2 \rangle \otimes \vert \beta_4 , \beta_5
\rangle
 \otimes \cdots \otimes \vert \beta_{3p-2}, \beta_{3p-1} \rangle
\end{equation}
where $\vert \beta_i, \beta_{i+1} \rangle$ $( i = 1, 4, \cdots ,
3p-2)$ are the $SU(3)$ coherent states defined in the previous
subsection. Explicitly, they are given by
\begin{equation}
\vert \beta_i, \beta_{i+1} \rangle = (1 + |\beta_i|^2 + |\beta_i
\beta_{i+1}|^2)^{-\frac{n}{2}}
\sum_{n_i=0}^{n}\sum_{n_{i+1}=0}^{n_i}
\frac{\beta_{i}^{n_i}\beta_{i+1}^{n_{i+1}}\sqrt{n!}}{\sqrt{(n -
n_i)!(n_i-n_{i+1})!n_{i+1}!}} \vert n-n_i , n_i-n_{i+1}, n_{i+1}
\rangle
\end{equation}
with the labeling parameters defined by
$$ \beta_i = i t_{i+1} \frac{r_i}{t_i} \quad {\rm and} \quad \beta_{i+1} = i \frac{r_{i+1}}{t_{i+1}},$$
are functions of transmission and reflection parameters.

This procedure to generate multipartite coherent states using a
quantum network of beam splitters can be extended to obtain
separable state involving the tensorial product of $SU(k+1)$
coherent states.

%Indeed, one can consider $mp-1$ beam splitters characterized by
%$$ \theta_{m} = \theta_{2m} = \cdots = \theta_{mp-m} = 0 \qquad {\rm for} \qquad p > 1 \qquad m \geq 2.$$
%In this case, it is easy to get  the action of the operator ${\cal
%U}_{mp-1}$ on the $mp$-particles of type $\vert n, 0,
%0,\cdots,0\rangle \otimes \vert n, 0, 0,\cdots,0\rangle \otimes
%\cdots \otimes \vert n, 0, 0,\cdots,0\rangle$ where $\vert n, 0,
%0,\cdots,0\rangle \in {\cal H}_1 \otimes {\cal H}_2 \otimes \cdots
%{\cal H}_m $. The resulting action is
% a tensorial product of $p$  coherent states associated with $SU(m)$
% symmetry similar to one giving by ({\ref{suncs}}) (modulo some obvious substitutions)

\section{ Entanglement of balanced superposition of $SU(2)$ multipartite coherent states}

The problem we raise in this section is the bipartite entanglement
of the balanced superposition of two multi-particle coherent states.
Before embarking on our study, it is worth making some remarks.
First, it should be noted that for the most of tasks of quantum
processing, one needs to generate such superpositions of
multipartite $SU(2)$ coherent states. In the next subsection, we
shall discuss their generation  via Kerr nonlinearity. Second, to
investigate the bipartite entanglement, we split the whole system in
two subsystems. Also, we shall focus on the multi-particle $SU(2)$
coherent states given by (\ref{su2}) generated via the scheme
described in the previous section. More precisely, we consider the
superposed states of the form
\begin{equation}
\vert \alpha;\alpha'\rangle \equiv {\cal N}_p [\vert \alpha_1
\rangle \otimes \vert \alpha_3 \rangle \otimes \cdots \otimes \vert
\alpha_{2p-1} \rangle + e^{i\theta} \vert \alpha'_1 \rangle \otimes
\vert \alpha'_3 \rangle \otimes \cdots \otimes \vert \alpha'_{2p-1}
\rangle], \label{alpha;alpha}
\end{equation}
to study the bipartite entanglement. It is clear that this
superposition is balanced or equally weighted.

%The first state is generated by beam splitters with transmission
%coefficients $t_i = 0$ for $i = 2, 4, \cdots, 2p-2$ and non
%vanishing coefficients $t_i$ when $i = 3, 5, \cdots, 2p-1$. The
%second state is obtained by the same technique but with different
%transmission coefficients ($t_i$ becomes $t_i'$).

The states $\vert \alpha_1 \rangle \otimes \vert \alpha_3 \rangle
\otimes \cdots \otimes \vert \alpha_{2p-1} \rangle$ and $\vert
\alpha'_1 \rangle \otimes \vert \alpha'_3 \rangle \otimes \cdots
\otimes \vert \alpha'_{2p-1} \rangle$ write as superpositions of
vectors states of $2p$ particles $\vert n_1, n_2, \cdots,
n_{2p-1},n_{2p}\rangle$. Remark that the analysis presented here can
be generalized to the multipartite $SU(k+1)$ coherent states in a
straightforward manner. The normalization constant in
(\ref{alpha;alpha}) is given by
$${\cal N}_p^{-2} = 2 ( 1 +  c_1 c_3\cdots c_{2p-1}~\cos\theta~)$$
where the quantities $c_i$ for $i = 1, 3,\cdots, 2p-1$  stand for
the overlapping of $SU(2)$ coherent states. They are given by
\begin{equation}
c_i \equiv c_i(\alpha_i, \alpha'_i) = \langle \alpha_i \vert
\alpha_i' \rangle
 = (t_it_i' + r_ir'_i)^n = \bigg[ \cos \frac{\theta_i - \theta'_i}{2}\bigg]^n
\label{overlap}
\end{equation}
in terms of the reflection-transmission coefficients.

We will employ the concurrence~\cite{Hill} as a measure of bipartite
entanglement for the state $\vert \alpha;\alpha'\rangle$. We recall
that for $\rho_{12}$ the density matrix for a pair of qubits~$1$
and~$2$ which may be pure or mixed, the concurrence is~\cite{Hill}
\begin{equation}
C_{12}=\max \left\{ \lambda _1-\lambda _2-\lambda _3-\lambda
_4,0\right\} \label{eq:c1}
\end{equation}
for~$\lambda_1\ge\lambda_2\ge\lambda_3\ge\lambda_4$ the square roots
of the eigenvalues of the "spin-flipped" density matrix
\begin{equation}
\varrho_{12}\equiv\rho_{12}(\sigma_y\otimes\sigma
_y)\rho_{12}^{\star}(\sigma_y\otimes \sigma_y), \label{eq:c2}
\end{equation}
where the star stands for complex conjugation in the basis $\{ \vert
00 \rangle, \vert 01 \rangle, \vert 10 \rangle, \vert 11 \rangle \}$
with the Pauli matrix is $\sigma_y = i \vert 1 \rangle \langle 0
\vert - i \vert 0 \rangle \langle 1 \vert$. For the special case of
pure state, one can show that the concurrence is $C_{12} = 2
\sqrt{{\rm det}\rho_1}$, where $\rho_1$ is the reduced density of
qubit 1 that is obtained by tracing out the second qubit. Nonzero
concurrence occurs if and only if qubits 1 and 2 are entangled.
Moreover, $C_{12}=0$ only for an unentangled state, and $C_{12}=1$
only for a maximally entangled state. This concurrence measure can
be used to study the bipartite entanglement in multipartite coherent
states for pure as well as mixed state as we shall explain below.

\subsection{Generation of balanced superposition of  multipartite $SU(2)$ coherent states}

As mentioned above, we shall discuss how to generate the balanced
superposition of multipartite $SU(2)$ coherent states. We first
focus on a  single beam splitter $(k = 1)$ with one port receives
$n$ photons and the vacuum at the other. We assume that a Kerr
medium is placed in the output of the beam splitter. The Kerr
interaction is described by the interaction Hamiltonian
$$H_{\rm Kerr} = \hbar \chi (a_1^+a_1^-)^2 $$
where $\chi$ is proportional to the third-order nonlinear
susceptibility of the medium. Many authors employ an extended
version where a linear term  in the photon number $a_1^+a_1^-$ is
added. This linear term is irrelevant for our task. The unitary
transformation associated with the Kerr interaction is
$$U_{\rm Kerr}(t) = \exp(-i t H_{\rm Kerr}/\hbar)$$
Clearly, the action of the beam splitter produces the $SU(2)$
coherent state $\vert \alpha_1 \rangle$ (see equation
(\ref{alphacs})) where $\alpha_1$ is related to the transmission and
reflection parameters of the device. In what follows we shall assume
that the time $t$ for the light to cross the Kerr medium is such
that $t = \pi/2\chi$. In this particular case, the action of the
operator $U_{\rm Kerr}(t)$ on the output state of the beam splitter
$\vert \alpha_1 \rangle$ gives
$$U_{\rm Kerr}(t) \vert \alpha_1 \rangle = \frac{1}{\sqrt{2}}(e^{-i\frac{\pi}{4}}\vert \alpha_1 \rangle
 + e^{+i\frac{\pi}{4}}\vert -\alpha_1 \rangle )$$

This can be generalized to generate some particular superposition of
multi-component $SU(2)$ coherent states. For this end, we use the
dynamical evolution of a tensorial product of $p$ $SU(2)$ coherent
states with respect the nonlinear Hamiltonian
$$H = \hbar \chi (a_1^+a_1^- + a_3^+a_3^- + \cdots + a_{2p}^+a_{2p}^-)^2.$$
This generalizes the single mode nonlinear Kerr hamiltonian. The
eigenvalues of this Hamiltonian are
$$H  \vert n_0 , n_1, n_2 , n_3, \cdots, n_{2p-1}, n_{2p} \rangle =
\hbar \chi (n_1+ n_3+\cdots+n_{2p})^2\vert n_0 , n_1, n_2 , n_3,
\cdots, n_{2p-1}, n_{2p} \rangle. $$ Here also at the time $t =
\frac{\pi}{2\chi}$, the multipartite coherent state $\vert \alpha_1
\rangle \otimes \vert \alpha_3 \rangle \otimes \cdots \otimes \vert
\alpha_{2p-1} \rangle$ evolves into the  state
\begin{equation}
e^{-itH/\hbar}\vert \alpha_1 \rangle \otimes \vert \alpha_3 \rangle
\otimes \cdots \otimes \vert \alpha_{2p-1} \rangle =
\frac{1}{\sqrt{2}}(e^{-i\frac{\pi}{4}} \vert \alpha_1 \rangle
\otimes \vert \alpha_3 \rangle \otimes \cdots \otimes \vert
\alpha_{2p-1} \rangle + e^{+i\frac{\pi}{4}} \vert -\alpha_1 \rangle
\otimes \vert -\alpha_3 \rangle \otimes \cdots \otimes \vert
-\alpha_{2p-1} \rangle),\label{balanced}
\end{equation}
which is a superposition of two multipartite $SU(2)$ coherent
states. It constitutes a special example of the balanced
multipartite $SU(2)$ coherent states given by (\ref{alpha;alpha}).

\subsection{Bipartite entanglement: Pure-state}

We now investigate the  degree of bipartite entanglement in
multipartite systems described by states of type $\vert
\alpha;\alpha'\rangle$ defined by (\ref{alpha;alpha}). Such states
belong to a $2p$ dimensional Hilbert space. As simple way to tackle
this issue, one  splits
 the entire system  into two subsystems as
follows
\begin{equation}
\vert \alpha;\alpha'\rangle \equiv {\cal N}_p [\vert \alpha)_q
\otimes \vert \alpha )_{p-q}  + e^{i\theta} \vert \alpha')_q \otimes
\vert \alpha')_{p-q}], \label{p,p-qstate}
\end{equation}
where
$$\vert \alpha)_q = \vert \alpha_1 \rangle \otimes \vert \alpha_3 \rangle \otimes \cdots \otimes \vert \alpha_{2q-1} \rangle
\qquad \vert \alpha )_{p-q} = \vert \alpha_{2q+1} \rangle \otimes
\vert \alpha_{2q+3} \rangle \otimes \cdots \otimes \vert
\alpha_{2p-1} \rangle$$ and
$$\vert \alpha')_q = \vert \alpha'_1 \rangle \otimes \vert \alpha'_3 \rangle \otimes \cdots \otimes \vert \alpha'_{2q-1} \rangle
\qquad \vert \alpha')_{p-q} = \vert \alpha'_{2q+1} \rangle \otimes
\vert \alpha'_{2q+3} \rangle \otimes \cdots \otimes \vert
\alpha'_{2p-1} \rangle$$ with $ 1 \leq q \leq p-1$. Thus, we
consider the first $2q$ systems  as the subsystem 1 containing $q$
$SU(2)$ coherent states and the other
 $2(p-q)$ systems as the subsystem 2 containing the remaining $p-q$ $SU(2)$ coherent states. Since the
two subsystems  in the bipartite state $\vert \alpha , \alpha'
\rangle$  are essentially two-state systems, we can characterize the
entanglement of bipartite state by the bipartite concurrence that we
denote by ${\cal C}_{q,p-q} \equiv {\cal C}_{(1,2\cdots
,2q)(2q+1,2q+2\cdots,2p)}$. It follows that one can apply the method
developed in \cite{Mann,Fu} to get the concurrence of a bipartite
system involving nonorthogonal states. For that end, one has to
write the multi-particle state  $\vert \alpha , \alpha' \rangle$ as
a state of two logical qubits using the orthogonal basis $\{ \vert
{\bf 0} \rangle_q , \vert {\bf 1} \rangle_q\}$ defined as
\begin{equation}
\vert {\bf 0} \rangle_q =  \vert \alpha)_q ,    \qquad \vert {\bf 1}
\rangle_q = \frac{\vert \alpha')_{q} - c_1c_3\cdots c_{2q-1} \vert
\alpha)_{q}}{\sqrt{1- c_1^2c_3^2\cdots c_{2q-1}^2}}, \label{base1}
\end{equation}
for the first subsystem. Similarly, we introduce for the second
subsystem the orthogonal basis $\{ \vert {\bf 0} \rangle_{p-q} ,
\vert {\bf 1} \rangle_{p-q}\}$ given by
\begin{equation}
\vert {\bf 0} \rangle_{p-q} =  \vert \alpha')_{p-q} ,   \qquad
\vert {\bf 1} \rangle_{p-q} =  \frac{\vert \alpha)_{p-q} -
c_{2q+1}c_{2q+3}\cdots c_{2p-1} \vert \alpha')_{p-q}}{\sqrt{1-
c_{2q+1}^2c_{2q+3}^2\cdots c_{2p-1}^2}}. \label{base2}
\end{equation}
Reporting (\ref{base1}) and (\ref{base2}) in (\ref{p,p-qstate}), we
obtain the expression of the pure state density $\vert
\alpha;\alpha'\rangle \langle \alpha ; \alpha' \vert$ in the basis
$\{ \vert {\bf 0} \rangle_{q} \otimes \vert {\bf 0} \rangle_{p-q} ,
 \vert {\bf 0} \rangle_{q} \otimes \vert {\bf 1} \rangle_{p-q} , \vert {\bf 1} \rangle_{q}
 \otimes \vert {\bf 0} \rangle_{p-q} , \vert {\bf 1} \rangle_{q} \otimes \vert {\bf 1} \rangle_{p-q}\}$,
from which one derives the reduced density matrix of the first
component of the system. Then it is straightforward  to check that
the concurrence is given by
\begin{equation}
{\cal C}_{q,p-q} =\frac{\sqrt{1- c_1^2c_3^2\cdots c_{2q-1}^2 }
\sqrt{1- c_{2q+1}^2c_{2q+3}^2\cdots c_{2p-1}^2 }}{ 1 +  c_1
c_3\cdots c_{2p-1}~\cos\theta}. \label{cq}
\end{equation}
Clearly, the separability condition of the state $\vert
\alpha;\alpha'\rangle $ is provided by the condition ${\cal
C}_{q,p-q} = 0$. This implies
\begin{equation}
c_1c_3\cdots c_{2q-1} = 1  \qquad {\rm or}  \qquad
c_{2q+1}c_{2q+3}\cdots c_{2p-1} = 1.
\end{equation}
Using the expression of the overlapping between two $SU(2)$ coherent
states (\ref{overlap}), it is simply verified that the first
condition is satisfied when $c_i = 1$ for each $i = 1, 3, \cdots,
2q-1$. This means that $\alpha_i = \alpha'_i$ or equivalently
$\theta_i = \theta'_i$ for all $i = 1,3, \cdots, 2q-1$. In this case
the state $\vert \alpha;\alpha'\rangle $ is unentangled.  Similarly,
the second condition is satisfied when $\alpha_i = \alpha'_i$ for $i
= 2q+1, 2q+3, \cdots, 2p-1$ and
it is easy to see that the state $\vert \alpha;\alpha'\rangle $ is separable. \\
The state $\vert \alpha;\alpha'\rangle$ is entangled when ${\cal
C}_{q,p-q} \neq 0$.
 From the expression of the concurrence (\ref{cq}), one can see that
maximally entangled states are obtained for $\theta = \pi$. The
minimal value of the concurrence is obtained for states with $\theta
= 0$ that we will call minimally entangled states.

At this level it is interesting to focus on some interesting
particular cases. In this respect, We consider the state $\vert
\alpha;\alpha'\rangle $ such that
\begin{equation}
\theta_1 - \theta'_1 = \theta_3 - \theta'_3 = \cdots = \theta_{2p-1}
- \theta'_{2p-1}. \label{conditiontheta}
\end{equation}
This condition, which can be  realized by some appropriate choice of
the variables labeling the $SU(2)$ coherent states, implies
\begin{equation}
c_1 = c_3 = \cdots = c_{2q-1} = c_{2q+1} = c_{2q+3}= \cdots =
c_{2p-1} = (\cos\varphi)^n \label{conditionc}
\end{equation}
where $ \varphi =( \theta_i - \theta'_i)/2$ takes the same value for
all $i = 1, 3, \cdots , 2p-1$. It follows that the concurrence
(\ref{cq}) takes the simple form
\begin{equation}
{\cal C}_{q,p-q} =\frac{\sqrt{1- c^{2nq} }\sqrt{1- c^{2n(p-q)} }}{ 1
+  c^{np}~\cos\theta}, \label{cqsimple}
\end{equation}
where $ c = \cos\varphi$.\\

For $p=2$, the only admissible value for $q$ is $q =1$. In this
case, the equation (\ref{cqsimple}) reads
\begin{equation}
{\cal C}_{1,1} =\frac{ 1- c^{2n} }{ 1 + c^{2n}~\cos\theta}.
\label{c11}
\end{equation}
It is clear that for $\theta = \pi$, the state is maximally
entangled. The concurrence ${\cal C}_{1,1}$  as function of the
overlapping $c$ and $\theta$ when the input state contains a single
photon ($n = 1$) is plotted in the figure 2.

\begin{center}
\includegraphics[width=4in]{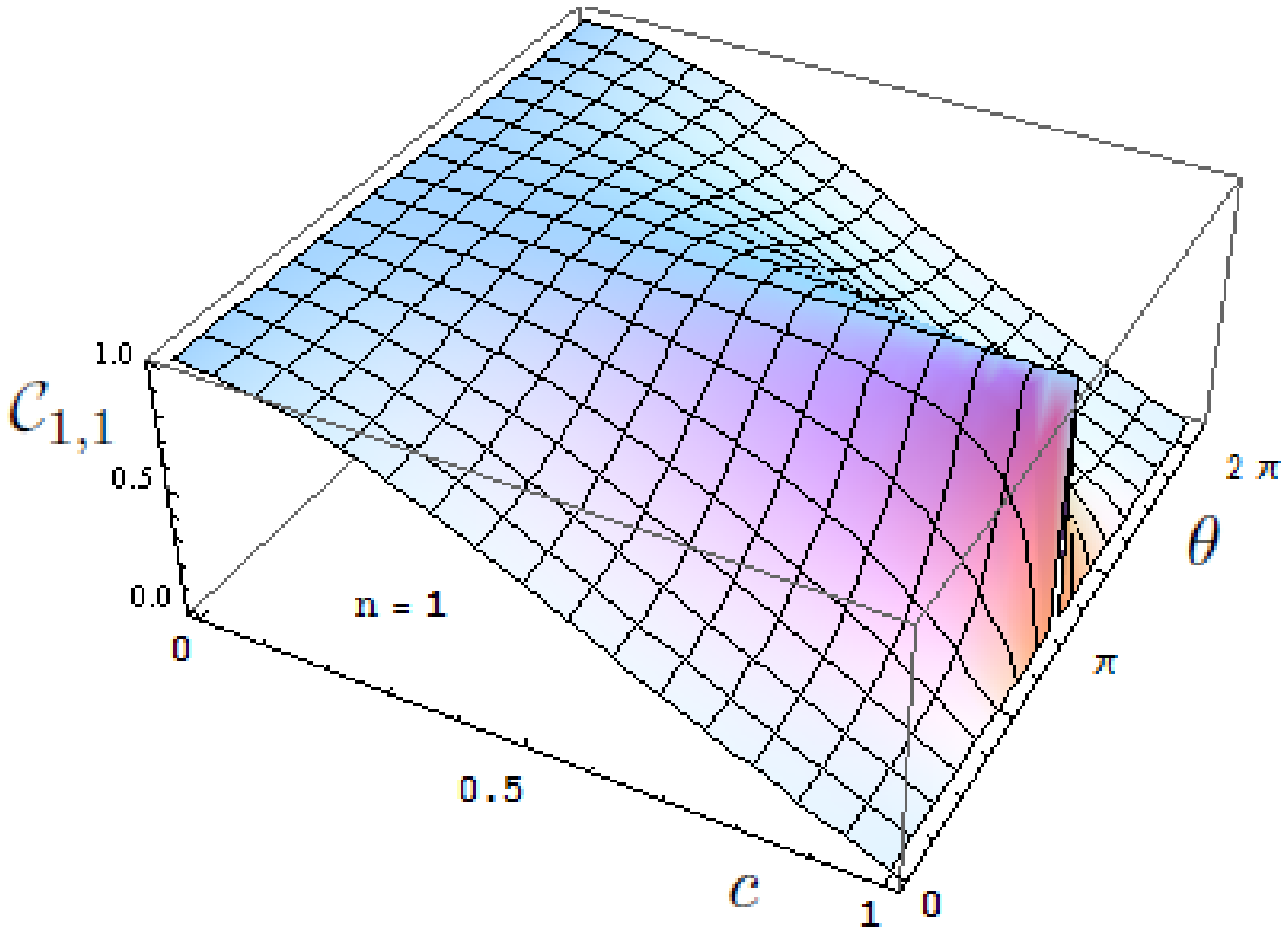}\\
Fig. 2: {\sf The concurrence ${\cal C}_{1,1}$  as function of $c$
and $\theta$ for $n=1$.}
\end{center}

For $\theta = 0$, the states  (\ref{p,p-qstate}), obtained by the
beam splitters satisfying the condition (\ref{conditiontheta}) or
equivalently (\ref{conditionc}), are completely symmetric. In this
case, we plot in figures 3 the concurrence ${\cal C}_{1,1}$  as
function of $\varphi = \arccos c $ for $\theta = 0$ and  different
input photon numbers $n$. It is remarkable that for $n$ increasing,
the concurrence increases quickly to reach the maximal entanglement
(${\cal C}_{1,1} = 1$).

\begin{center}
\includegraphics[width=4in]{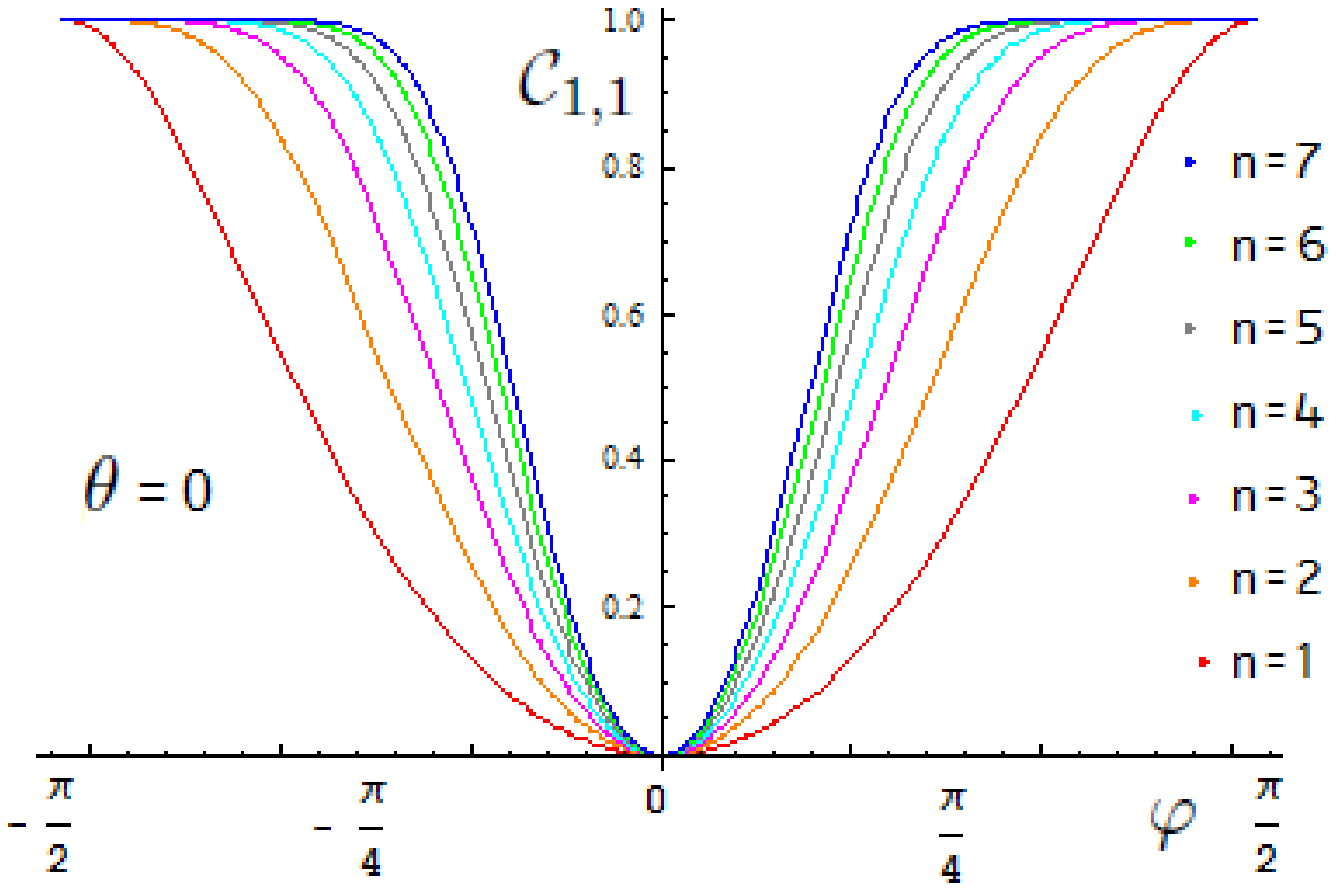}\\
Fig. 3: {\sf The concurrence ${\cal C}_{1,1}$  as function of
$\varphi = \arccos c $ for $\theta = 0$ and  different input photon
numbers $n$.}
\end{center}

Finally, we plotted also the ${\cal C}_{1,1}$  as function of
$\varphi = \arccos c $ for $\theta = \frac{\pi}{2}$ and  different
input photon numbers $n$. This is given by the figure 4. In this
case, as it can be be seen from the figure, the concurrence
increases with increasing number of input photons $n$.

\begin{center}
\includegraphics[width=4in]{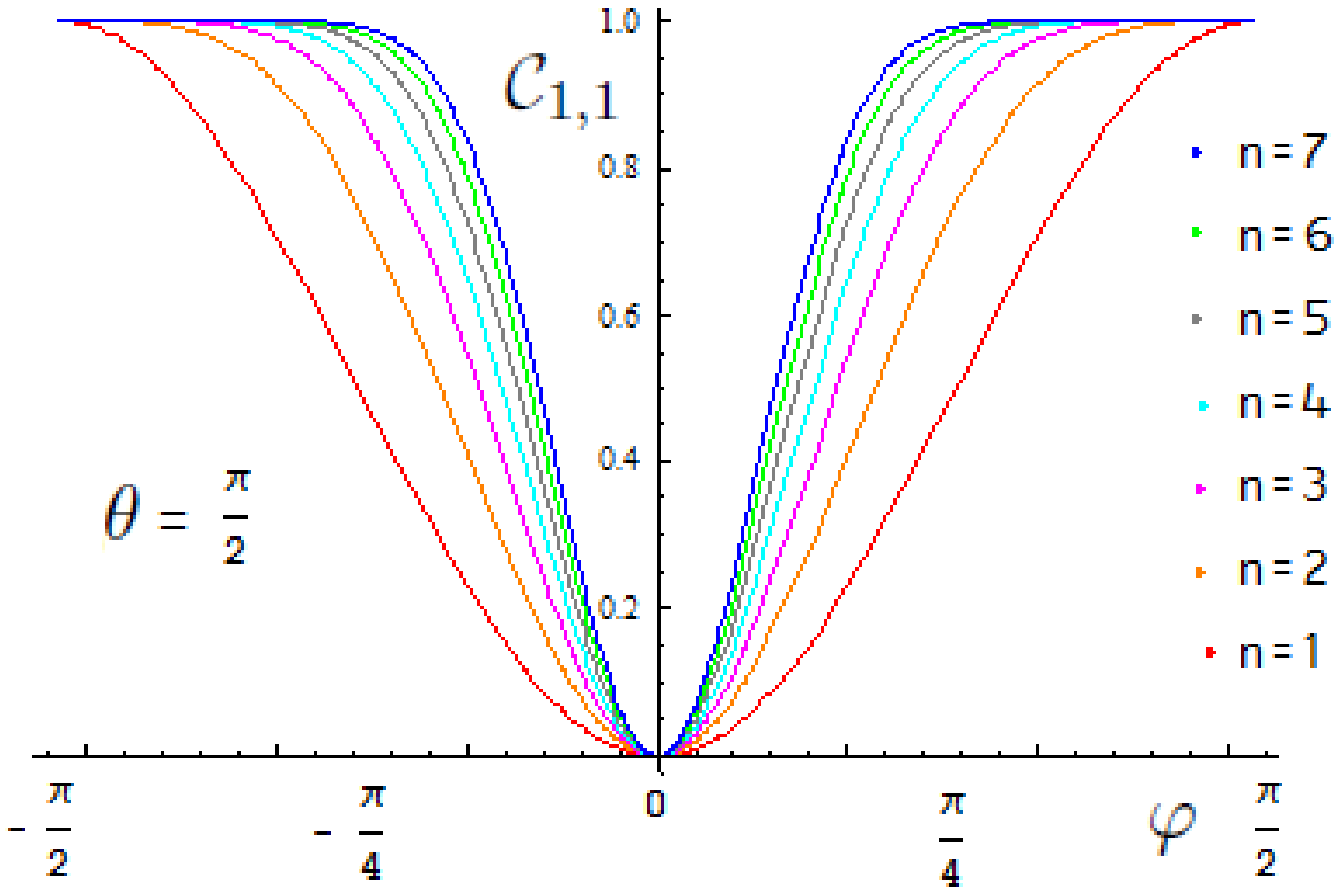}\\
Fig. 4: {\sf The concurrence ${\cal C}_{1,1}$  as function of
$\varphi = \arccos c $ for $\theta = \frac{\pi}{2}$ and  different
input photon numbers $n$.}
\end{center}

As another illustration of the above analysis, we also consider the
situation where $p=3$. In this case, one can divide the system such
that $q = 1$ or $q=2$. In this case, due to the symmetry property
${\cal C}_{q,p-q} = {\cal C}_{p-q,q}$, one has
\begin{equation}
{\cal C}_{1,2} = {\cal C}_{2,1} = \frac{\sqrt{1- c^{2n} }\sqrt{1-
c^{4n} }}{ 1 +  c^{3n}~\cos\theta}
\end{equation}
In figures 5, 6 and 7, we plot the concurrence ${\cal C}_{1,2}$ as
function of $\varphi = \arccos c $ and $\theta $ for $n = 1$, $n =
5$ and $ n = 10 $.

\begin{center}
\includegraphics[width=4in]{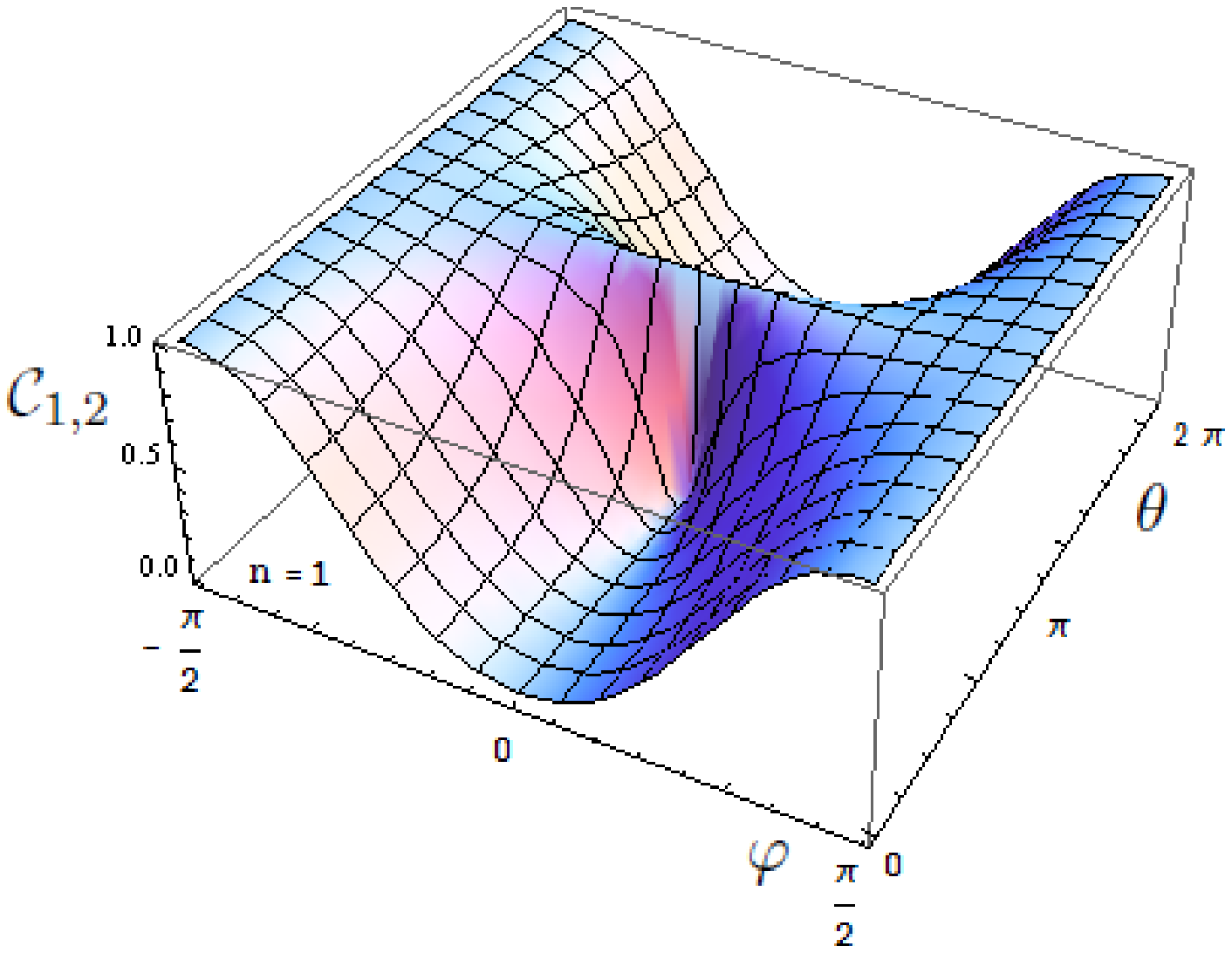}\\
Fig. 5: {\sf The Concurrence ${\cal C}_{1,2}$  as function of
$\varphi$ and $\theta$ for $n=1$.}
\end{center}

\begin{center}
\includegraphics[width=4in]{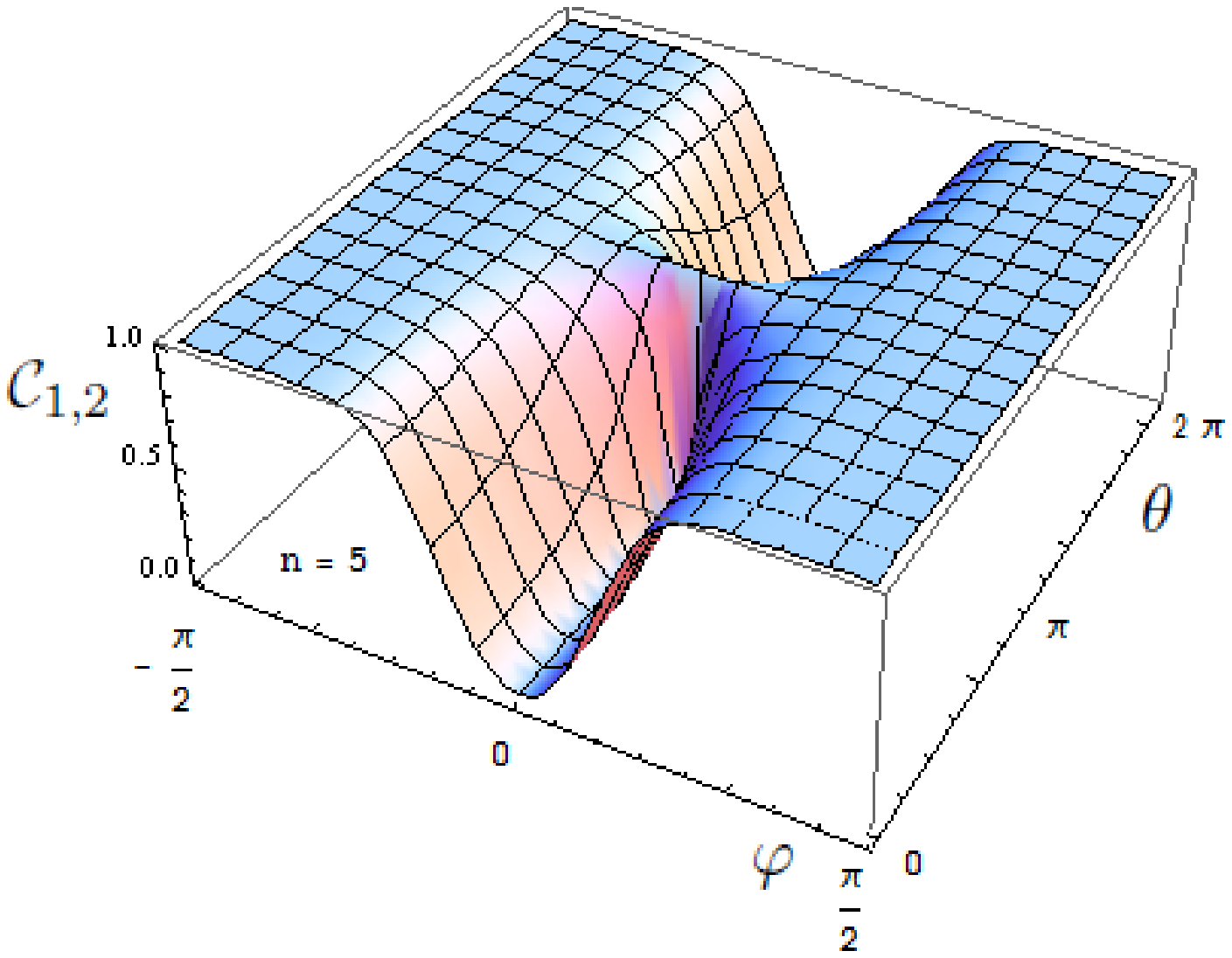}\\
Fig. 6: {\sf  The Concurrence ${\cal C}_{1,2}$  as function of
$\varphi$ and $\theta$ for $n=5$.}
\end{center}

\begin{center}
\includegraphics[width=4in]{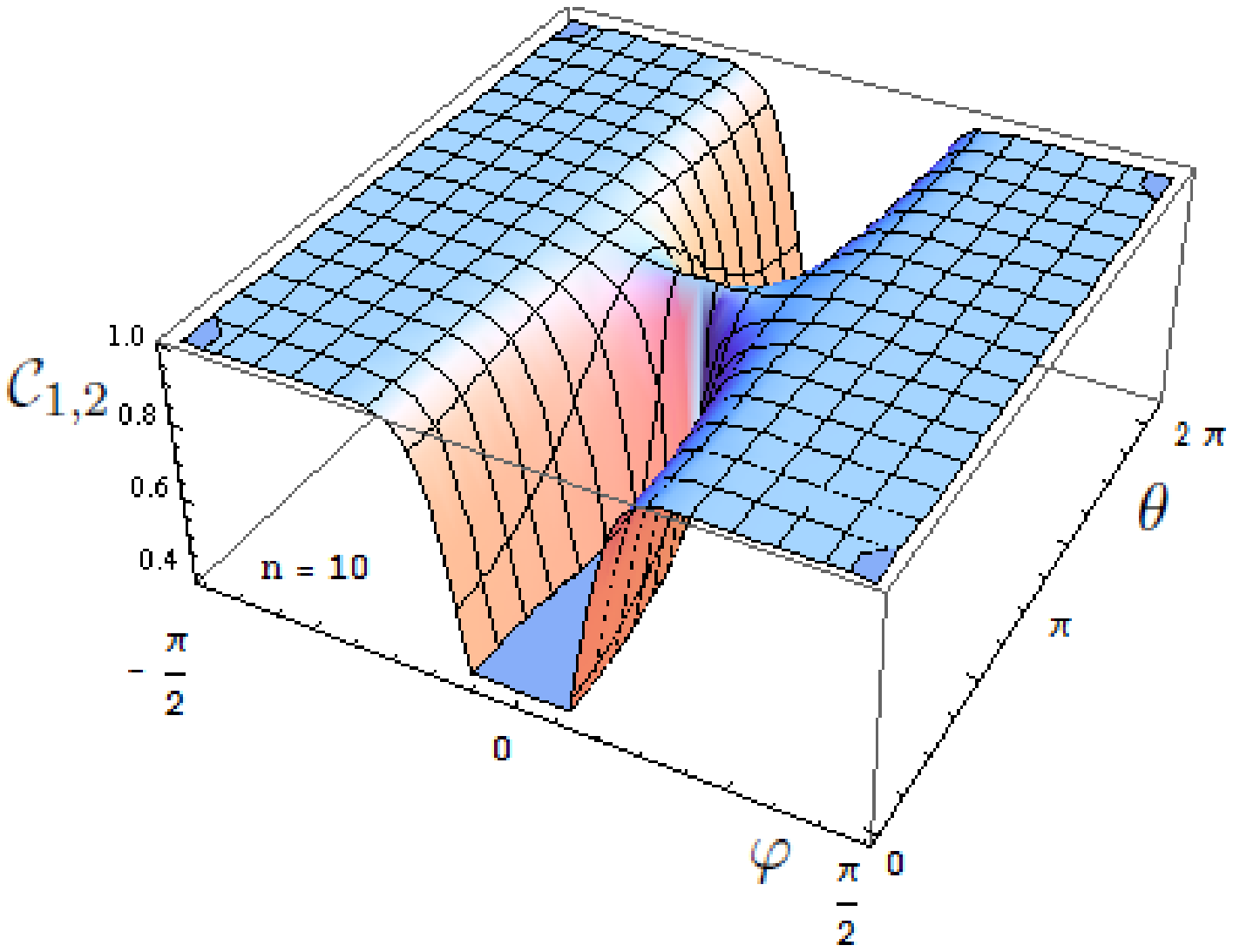}\\
Fig. 7: {\sf The Concurrence ${\cal C}_{1,2}$  as function of
$\varphi$ and $\theta$ for $n=10$.}
\end{center}

\subsection{Bipartite entanglement: Mixed-state}
Another scheme to deal with the bipartite entanglement of the state
$\vert \alpha;\alpha'\rangle$ can be achieved as follows. As $\vert
\alpha;\alpha'\rangle$ is defined on a $2p$ dimensional Hilbert
space, a bipartite state can be defined by tracing out $(2p-4)$
harmonic oscillator degrees of freedom
\begin{equation}
\rho_{(1234)} = {\rm Tr}_{5,\cdots ,2p}\vert \alpha ;\alpha'\rangle
\langle \alpha ;\alpha'\vert.
\end{equation}
It results that the reduced density matrix $\rho_{13} \equiv
\rho_{(1234)}$ can be cast in the following form
\begin{equation}
\rho_{13} = {\cal N}_p^2 [\vert \alpha_1 ,\alpha_3\rangle \langle
\alpha_1,\alpha_3\vert + c_5c_7\cdots c_{2p-1} e^{-i\theta}\vert
\alpha_1 ,\alpha_3\rangle  \langle \alpha'_1,\alpha'_3\vert +
c_5c_7\cdots c_{2p-1} e^{+i\theta} \vert \alpha'_1 ,\alpha'_3\rangle
\langle \alpha_1,\alpha_3\vert + \vert \alpha'_1 ,\alpha'_3\rangle
\langle \alpha'_1,\alpha'_3\vert ]. \label{eq:mix}
\end{equation}
It only involves the $SU(2)$ coherent states labeled by the
parameters $\alpha_1$ and $\alpha_3$ and it is indeed shared by two
subsystems. The first (resp. second) is described by the bosonic
degrees of freedom indexed by 1 and 2 (resp. 3 and 4).

Note that this is not the only way to introduce a bipartite density
matrix and  there are $p(p-1)/2$ different density matrices
$\rho_{kl}$. However, for our state $\vert \alpha;\alpha'\rangle$,
all particles are equally entangled with each other and all the
reduced density matrices $\rho_{kl}$ are identical.
Therefore, it is sufficient to consider $\rho_{13}$ and to generalize from this case.\\

To simplify our purpose let us assume that the state $\vert
\alpha;\alpha'\rangle$ is prepared such that
 $\alpha_1 = \alpha_3'$ and $\alpha_3 = \alpha_1'$. To obtain the concurrence (\ref{eq:c1}), one
has to diagonalize the density matrix (\ref{eq:c2}). For this, we
choose an orthogonal basis $\{\vert {\bf 0}\rangle ,\vert {\bf
1}\rangle \}$ defined as
\begin{equation}
\vert{\bf 0}\rangle\equiv\vert\alpha_1 \rangle~,\qquad \vert{\bf
1}\rangle\equiv(\vert \alpha_3 \rangle - \langle \alpha_1 \vert
\alpha_3 \rangle \vert {\bf 0}\rangle)/\sqrt{ 1 - \langle \alpha_1
\vert \alpha_3 \rangle^2}.
\end{equation}
It follows that substituting
\begin{equation}
\vert\alpha_1 \rangle~ = \vert{\bf 0}\rangle \qquad \vert \alpha_3
\rangle= \sqrt{ 1 - \langle \alpha_1 \vert \alpha_3 \rangle^2}
\vert{\bf 1}\rangle +  \langle \alpha_1 \vert \alpha_3 \rangle
\vert{\bf 0}\rangle
\end{equation}
into Eq.~(\ref{eq:mix}), one has  the  density matrix $\rho_{13}$ in
the  basis $\{\vert{\bf 00}\rangle ,\vert{\bf 01}\rangle ,\vert{\bf
10}\rangle ,\vert{\bf 11}\rangle \}$ and subsequently one obtains
the "spin-flipped" density matrix $\varrho_{13}$ (defined as in
(\ref{eq:c2})). The square roots of eigenvalues of $\varrho_{13}$
are
\begin{equation}
\lambda_1 ={\cal N}_p^2 (1 - \langle \alpha_1 \vert \alpha_3
\rangle^2) (1 + c_5c_7\cdots c_{2p-1}), \lambda_2 ={\cal N}_p^2 (1 -
\langle \alpha_1 \vert \alpha_3 \rangle^2) (1 - c_5c_7\cdots
c_{2p-1}), \lambda_3 =\lambda_4 = 0,
\end{equation}
and the concurrence is thus  given by
\begin{equation}
{\cal C}_{13} (p) \equiv {\cal C}_{(12)(34)} (p) = \frac{(1 -
\langle \alpha_1 \vert \alpha_3 \rangle^2) c_5c_7\cdots c_{2p-1}}{1
+  c_1 c_3\cdots c_{2p-1}~\cos\theta~}. \label{c13}
\end{equation}
Remark that, as we assumed $\alpha_1 = \alpha_3'$ and $\alpha_3 = \alpha_1'$, we have $ c_1 = c_3 = \langle \alpha_1 \vert \alpha_3 \rangle $.\\

In the particular situation where  $p = 2$, the concurrence
(\ref{c13}) reads as
\begin{equation}
{\cal C}_{13}(p=2) =\frac{1 - \langle \alpha_1 \vert \alpha_3
\rangle^2}{1 +  \langle \alpha_1 \vert \alpha_3
\rangle^2~\cos\theta~}. \label{plot}
\end{equation}
This quantity depends on the number $n$ of photons passing trough
the beam splitters network (see equation (\ref{conditionc})), the
difference phase orientations of the beam splitters 1 and 3, i.e.
$\theta_1 - \theta_3$, and the parameter $\theta$. Indeed, the
concurrence (\ref{plot}) can also be written in the form
\begin{equation}
{\cal C}_{13} (p=2) =\frac{1 - (\cos\theta_{13})^{2n}}{1 +
(\cos\theta_{13})^{2n}~\cos\theta~},
\end{equation}
where $\theta_{13} = (\theta_1 - \theta_3)/2$. It is clear that for
$\theta_{13} = 0$ or equivalently $\alpha_1 = \alpha_3$, the
concurrence vanishes. For fixed $\theta_{13}$, the the concurrence
${\cal C}_{13} (p=2)$ is maximal for $\theta = \pi$. It is also
remarkable that there is a formal similarity between the expression
${\cal C}_{13} (p=2)$ and the concurrence ${\cal C}_{1,1}$ given by
the equation (\ref{c11}) (modulo the substitution $c = \cos \varphi
\to \cos\theta_{13}$). Hence,  the behavior of the concurrence
${\cal C}_{13} (p=2)$ as function of $\cos\theta_{13}$, $\theta$ and
the photon number $n$, initially injected in the network to generate
the coherent states, is similar to one exhibited by the concurrence
${\cal C}_{1,1}$ (see the figures 2, 3 and 4).

\section{\bf Summary}
In view of the first part of the paper we can state that any
$SU(k+1)$ coherent state in the Perelomov sense can be generated by
using a network composed by $k$ beam splitters. The continuous
parameters labeling such states are expressed in term of the
reflection-transmission parameters of the considered beam splitters.
Based on this result, we have discussed the bipartite entanglement
of a balanced superposition of multipartite coherent states. We
particularly investigated the bipartite entanglement of
multi-component $SU(2)$ coherent states. We discussed the pure as
well as the mixed states cases. We gave the evolution of the
concurrence as function of the number of photons initially injected
in the network, the transmission parameters of the beam splitters.
This is corroborated by numerical analysis in some particular
situations. This shows clearly that the generation of multipartite
entangled coherent can simply be achieved using a network of beam
splitters.

\vspace{1cm}

{\bf Acknowledgments}: \\
MD would like to express his thanks to Max Planck Institute for
Physics of Complex Systems (Dresden-Germany)
where this work was done.\\

%\newpage


\begin{thebibliography}{99}


\bibitem{Ben1} C.H.  Bennett, G. Brassard, C, Cr\'epeau, R. Jozsa, A. Peres and W.K. Wootters, Phys. Rev. Lett. {\bf 70}  (1993) 1895.

\bibitem{Ben2} C.H. Bennett and S.J. Wiesner, Phys. Rev. Lett. {\bf 69}
 (1992) 2881.

\bibitem{Eckert} A.K. Ekert, Phys. Rev. Lett. {\bf 67}  (1991) 661.

\bibitem{Fuchs}  C.A. Fuchs, Phys. Rev. Lett. {\bf 79}  (1997) 1162.

\bibitem{Rausschendorf}  R. Rausschendorf and H. Briegel, {\tt quant-ph/0010033}.

\bibitem{Gottesman}  D. Gottesman and I. Chuang, Nature  {\bf 402} (1999) (6760) 390.

\bibitem{Rungta}  P. Rungta, V. Buzek, C.M. Caves, M. Hillery, and G.J. Milburn, Phys. Rev. A {\bf 64} (2001) 042315.

\bibitem{Ben3}  C.H. Bennett, D.P. DiVincenzo, J. Smolin, and W.K.
Wootters, Phys. Rev. {\bf A 54} (1996) 3824.

\bibitem{Wootters}  W.K. Wootters, Phys. Rev. Lett. {\bf 80} (1998) 2245.

\bibitem{Coffman}  V. Coffman, J. Kundu, and W.K. Wootters, Phys. Rev.
A {\bf 61} (2000) 052306.

\bibitem{Kwiat} P.G. Kwiat, S. Barraza-Lopez, A. Stefanov and N. Gisin, Nature {\bf 409} (2001) 1014.

\bibitem{Tan}  S.M. Tan, D.F. Walls and M.J. Collett, Phys. Rev. Lett. {\bf 66} (1991) 252.

\bibitem{Sanders1}  B.C. Sanders, Phys. Rev.  {\bf A 45} (1992) 6811.

\bibitem{Sanders2}  B.C. Sanders, K.S. Lee and M.S. Kim, Phys. Rev. {\bf A 52} (1995) 735.

\bibitem{Paris}  M.G.A. Paris, Phys. Rev. {\bf A 59} (1999) 1615.

\bibitem{Kim}  M.S. Kim, W. Son, V.Buzek  and  P.L. Knight, Phys. Rev. {\bf A 65} (2002) 032323.

\bibitem{Mann}  A. Mann, B. C. Sanders, and W. J. Munro, Phys. Rev. A {\bf 51}, 989 (1995).

\bibitem{Wang}  X. Wang, B.C. Sanders  and S.H. Pan, J. Phys. A:
Math. Gen. {\bf 33}  (2000) 7451.

\bibitem{Fu}  H. Fu, X. Wang and A. I. Solomon, Phys. Lett. A, {\bf 291} (2001) 73.

\bibitem{van} S.J. van Enk, O. Hirota, Phys. Rev. {\bf A 64} (2001) 022313.

\bibitem{Jeong1} H. Jeong, M.S. Kim, J. Lee, Phys. Rev. {\bf A 64} (2001) 052308.

\bibitem{Jeong2}  H. Jeong, M.S. Kim, Phys. Rev. {\bf A 65} (2002) 042305.

\bibitem{Yang} M. Yang, Z.-L. Cao, Physica {\bf A 366} (2006) 243.

\bibitem{Sangouard} N. Sangouard, C. Simon, N. Gisin, J. Laurat, R. Tualle-Brouri and P. Grangier, J. Opt. Soc. Am. {\bf B 27} (2010) A137.

\bibitem{Markham}  D. Markham  and V. Vedral,  Phys. Rev. {\bf A 67}  (2003) 042113.

\bibitem{Gerry1}  C.C.  Gerry and A. Benmoussa, Phys. Rev. {\bf A 71} (2005) 062319.

\bibitem{Klauder}  J.R. Klauder and B.S. Skagerstam, Coherent states-Applications in Physics and Mathematical Physics (World Scientific,Singapore,1985).

\bibitem{Perelomov}  A. Perelomov, Generalized Coherent States and their Applications,Texts and Monographs in Physics, (Spinger-Verlag,1986).

\bibitem{Gerry2} C.C. Gerry, Phys. Rev. {\bf A 59} (1999) 4095.

\bibitem{Luis}  A. Luis, Phys. Rev. {\bf A 64} (2001) 054102.

\bibitem{Wang1}  X. Wang and B.C. Sanders, Phys. Rev. {\bf A 65} (2002) 012303.

\bibitem{Wang2}  X. Wang, J. Phys. A: Math. Gen. {\bf 35} (2002) 165.


\bibitem{Hill}  S. Hill and W. K. Wootters, Phys. Rev. Lett. {\bf 78}
(1997) 5022; W. K. Wootters, Phys. Rev. Lett. {\bf 80} (1998) 2245.

\bibitem{Van}  P. van Loock and S.L. Braunstein, Phys. Rev. Lett. {\bf 84} (2000) 3482.

\bibitem{Daoud1} M. Daoud, Phys. Lett. {\bf A 329} (2004) 318.

\bibitem{Brattke} S. Brattke, B.T.H. Varcoe and H. Walther, Phys.
Rev. Lett. {\bf 86} (2001) 3534.

\bibitem{Hofheinz} M. Hofheinz, E.M. Weig, M. Ansmann, R.C.
Bialczak, E. Lucero, M. Neeley, A.D. O'Connell, H. Wang, J.M.
Martinis and A.N. Cleland, Nature {\bf 454} (2008) 310.


%\bibitem{Daoud} M. Daoud and  M.R.  Kibler, J. Phys. A: Math. Theor. {\bf 43} (2010) 115303.


%\bibitem{Ma} Z. Ma and M. Bao, Phys. Rev {\bf A 82} (2010) 034305.






%%%%%%%%%%%%%%%%%%%%%%%%%%%%%%%%%%%%%%%%%%%%%%%%%%%%%%%%%%%%%%%%%%%%%%%%%%%%%%%%555




\end{thebibliography}
\end{document}